# Non-Orthogonal Multi-band CAP for Highly Spectrally Efficient VLC Systems

Paul Anthony Haigh[1*], Petr Chvojka[2*], Z. Ghassemlooy[3], Stanislav Zvanovec[2] and Izzat Darwazeh[1]

[1]Communications and Information Systems Group, University College London, London, WC1E 6BT, UK
[2]Department of Electromagnetic Field, Czech Technical University in Prague, Technicka 2, 16627, Prague, Czech Republic
[3]Optical Communications Research Group, Northumbria University, Newcastle-upon-Tyne, NE1 8ST, UK
{p.haigh; i.darwazeh}@ucl.ac.uk, {chvojpe8; xzvanove}@fel.cvut.cz, z.ghassemlooy@northumbria.ac.uk
[*]*These authors contributed equally*

*Abstract*—In this work we propose and experimentally demonstrate a novel non-orthogonal multi-band carrier-less amplitude and phase (NM-CAP) scheme for bandlimited visible light communication systems in order to increase the spectral efficiency. We show that a bandwidth saving up to 30% can be achieved thus resulting in 44% improvement in the measured spectral efficiency with no further bit error rate performance degradation compared to the traditional *m*-CAP scheme. We also show that higher order systems can provide higher bandwidth compression than low order systems. Furthermore, with no additional functional blocks at the transmitter or the receiver the proposed scheme introduces no extra computational complexity.

## I. INTRODUCTION

In recent years, visible light communications (VLC) has attracted enormous attention within the academic and industrial communities worldwide [1]. VLC mostly utilises white light-emitting diodes (LEDs) to support simultaneous data transmission and illumination in a home/office environment, and hence, is a promising solution for the last-meter connectivity in future communication networks (i.e., 5G) [2]. Due to the limited modulation bandwidth (i.e. a few MHz) of the most commonly available LEDs [3, 4], the vast majority of research activities have focused on implementation of techniques to demonstrate the potential of VLC links at high data rates (i.e. from hundreds of Mb/s to Gb/s). The techniques used are often based on multi-carrier modulation schemes including orthogonal frequency division multiplexing (OFDM) or carrier-less amplitude and phase (CAP) modulation [5, 6]. Generally, the research community aims to achieve higher throughput without considering the specific spectral efficiencies of VLC systems, which is a key performance metric rarely reported as part of the link bit rate performances evaluation. Recently, a number of methods have been proposed to increase spectral usage, such as fast OFDM (FOFDM) as originally proposed in [7], spectrally efficient FDM (SEFDM) [8], dense OFDM (DOFDM) [9] or faster than Nyquist (FTN) signalling [10]. FOFDM maintains the spectral efficiency of an OFDM system using Hermitian symmetry, albeit occupying half the bandwidth by reducing the minimum frequency spacing between subcarriers to half that of the symbol rate, i.e., $1/(2T)$. Unfortunately, FOFDM is limited only to one-dimensional modulation schemes (e.g., binary phase shift keying (BPSK) or pulse amplitude modulation (PAM)), hence the constant spectral efficiency in comparison to OFDM. On the other hand, SEFDM supports two-dimensional modulation schemes, with non-orthogonal subcarrier spacing, which increases the spectral efficiency at the cost of self-induced inter-carrier interference (ICI) between subcarriers. In [8] it was experimentally demonstrated that up to 25% bandwidth can be saved by SEFDM in comparison to OFDM with negligible bit error rate (BER) performance degradation. Contrary to OFDM-based techniques, the FTN method is implemented in the time domain and operates on the basis of non-orthogonal filter design. However, akin to SEFDM, the improvement in the achieved spectral efficiency comes at the cost of the increased transmitter and receiver complexities [10].

Multi-band CAP (*m*-CAP) modulation scheme has been experimentally demonstrated as a potential alternative candidate for highly bandlimited VLC systems [11, 12]. Unlike OFDM, *m*-CAP utilizes band-pass finite impulse response (FIR) filters to transmit data streams in orthogonally spaced subcarriers as opposed to the inverse fast Fourier transform (IFFT). The FIR filters, which also generate the carrier frequencies, are deployed at both the transmitter and receiver. These are crucial to maintain the required overall system performance and account for most of the system complexity as demonstrated in [12]. By optimizing the number of filters used and their parameters (i.e. the number of taps and the roll-off factor $\beta$, which determines the excess bandwidth) the required system throughput and spectral efficiency can be improved [12]. In [13, 14], another approach for improving the VLC link capacity/spectral efficiency was introduced by applying unequally spaced subcarriers. The most recent works have reported transmission speed and spectral efficiency up to ~250 Mb/s [15] and ~6 b/s/Hz [12], respectively, thus providing sufficient headroom for further improvement in VLC systems as reported based on numerical simulations in [16].

Hence, in this work we propose, for the first time, and experimentally verify a novel non-orthogonal scheme termed non-orthogonal *m*-CAP (NM-CAP). This system advantageously improves the spectral efficiency in VLC systems. The proposed novel scheme is based on modification of the filter carrier frequencies to force compression of subcarrier spacing, below orthogonality, and hence cause overlapping in the adjacent filter passbands. We analyse the

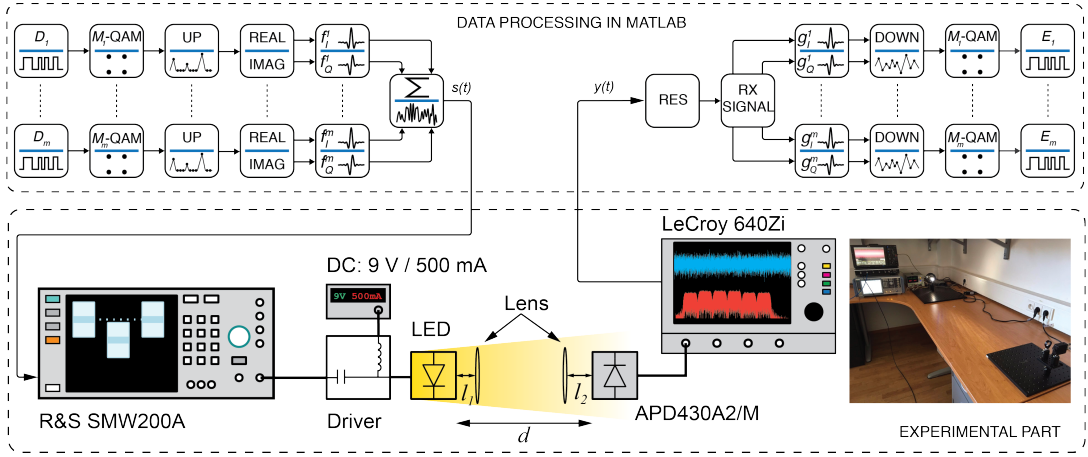

Fig. 1 The schematic block diagram of the experimental NM-CAP VLC system. 'UP', 'RES' and 'DOWN' blocks refer to up-sampling, resampling and down-sampling, respectively

system BER performance for a bandwidth compression up to 50% for a range of excess bandwidth parameter $\beta = \{0.1, 0.2, 0.3\}$ and number of subcarriers $m = \{2, 10\}$. These values are chosen because the minimum number of subcarriers that can be used is 2 due to the requirement for multiple bands, while $m = 10$ is selected because the literature has demonstrated that the for setting $m > 10$, only limited improvement in spectral efficiency or data rate [15] is attained. The results show that a bandwidth saving up to 30% can be achieved with no further degradation in BER performance. Moreover, there is no need for modification of the receiver structure and hence, no increase in complexity in comparison to traditional $m$-CAP

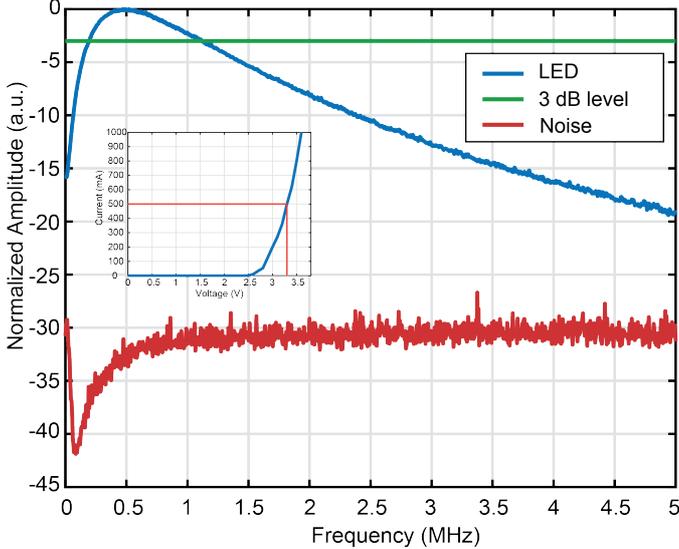

Fig. 2 The measured LED and noise frequency responses with the highlighted 3 dB level. The band-pass feature of the LED magnitude response is cause by the bias-tee acting as a high-pass filter and such, causing a cut-on effect at around 250 kHz. The measured cut-off frequency is around 1.25 MHz, giving an effective bandwidth of around 1 MHz. Inset, the measured I-V curve of the LED and drive circuit.

The rest of the paper is organized as follows. The system setup and the principles of the NM-CAP scheme are described in Section II. The results are discussed in Section III, and finally the conclusions are given in Section IV.

## II. EXPERIMENTAL SETUP

A simplified block diagram of the experimental setup for the proposed NM-CAP scheme is illustrated in Fig. 1. A pseudorandom binary sequence (PRBS) $D_m$ of length $2^{15}-1$ is repeated for $10^6$ bits for each subcarrier. The individual bits are mapped onto an $M$-ary quadrature amplitude modulation ($M$-QAM), where $M$ is the order of QAM and was set to 4 in this work. The symbol sequences are up-sampled according to the number of samples/symbol $n_s = \lceil 2m \cdot (1 + \beta) \rceil$ where $\lceil \cdot \rceil$ is the ceiling function. The signal is then split into its real and imaginary parts (i.e., or in-phase ($I$) and quadrature ($Q$)) and passed through the square root raised cosine (SRRC) pulse shaping filters. The impulse responses of the transmit filters forms a Hilbert pair (i.e., they are orthogonal in the time domain with a 90° phase shift) and are given as a product of the SRRC filter impulse response and the cosine and sine waves for the $I$ and $Q$ signal, respectively, which are given respectively as [12, 15]:

$$f_I^m(t) = \begin{bmatrix} \dfrac{sin[\gamma(1-\beta)] + 4\beta \dfrac{t}{T_s} cos[\gamma\delta]}{\gamma \left[1 - \left(4\beta \dfrac{t}{T_s}\right)^2\right]} \\ \times cos[\gamma(2m-1)\delta] \end{bmatrix} \quad (1)$$

and

$$f_Q^m(t) = \begin{bmatrix} \dfrac{sin[\gamma(1-\beta)] + 4\beta \dfrac{t}{T_s} cos[\gamma\delta]}{\gamma \left[1 - \left(4\beta \dfrac{t}{T_s}\right)^2\right]} \\ \times sin[\gamma(2m-1)\delta] \end{bmatrix} \quad (2)$$

where $T_s$ is the symbol duration, $\gamma = \pi t/T_s$ and $\delta = 1+\beta$. The frequencies of carriers, generated by the pulse shaping transmit filters, will be given by:

$$f_c^n = \frac{2n-1}{2m} B(1-\alpha) \quad (3)$$

where $n$ is the subcarrier number, $B$ is the total signal bandwidth and $\alpha$ is defined here as a bandwidth compression factor.

In the conventional $m$-CAP scheme, $\alpha$ is set to 0, thus maintaining the orthogonality between subcarriers. In the proposed NM-CAP system, we shift the carrier frequencies towards lower values, compressing by an overall factor $\alpha$, purposely overlapping the subcarriers and breaking their orthogonality. The outputs of the filters are real-valued, and are summed to form the discrete time domain signal as given by [12]:

$$s(t) = \sqrt{2} \sum_{n=1}^{m} \left( s_I^n(t) \otimes f_I^n(t) - s_Q^n(t) \otimes f_Q^n(t) \right) \quad (4)$$

where $\otimes$ represents the time domain convolution and $s_I^n(t)$ and $s_Q^n(t)$ are in phase and quadrature components of the $M$-QAM symbols for the $n^{th}$ subcarrier, respectively. Note that the discrete signal $s(t)$ is sampled at the sampling frequency given by:

$$f_s = R_s n_s \frac{1}{m} \quad (5)$$

where $R_s$ is the signal baud rate.

The generated discrete signal $s(t)$ is loaded into a Rohde & Schwarz SMW200A vector signal generator, which in turn feeds the LED driving circuit prior to intensity modulation of the light source. The light source used was a commercially available LED (OSRAM Golden Dragon) and the measured frequency response showing a 3 dB modulation bandwidth of ~1.2 MHz is illustrated in Fig. 2. Note that, we used a bias current of 500 mA to ensure the intensity modulation within the LED linear region, see inset in Fig. 2. The modulated light was transmitted over a range $d = 2$ m, which is typical for indoor VLC links). At the Rx, a low noise avalanche photodetector (APD) (Thorlabs APD430A2/M) based optical receiver with inbuilt transimpedance amplifier was used to regenerate the electrical signal. Two 25.4 mm biconvex lenses were placed at distances of $l_1 = 25$ mm and $l_2 = 35$ mm from the LED and the PD, respectively, to focus and collimate the emitted and received light beams. The regenerated signal was captured using a LeCroy WaveRunner Z640i real time oscilloscope with a sampling rate of 50 MS/s for further offline processing in MATLAB.

The signal processing adopted to recover the data information is the same as in the traditional $m$-CAP scheme [15]. Following resampling (denoted 'RES' in Fig. 1), the signal is passed through the time-reversed filters matched to the transmit filters as $g_I^m(t) = f_I^m(-t)$ and $g_Q^m(t) = f_Q^m(-t)$ for the real and imaginary parts of the signal, respectively, see Fig. 1. The signal is down-sampled and demodulated to recover the received $M$-QAM symbols for BER estimation. The resulting BER performance is determined by comparing the transmitted and received bits in a bit-by-bit manner, and we set the BER floor to $10^{-4}$.

III. RESULTS AND DISCUSSION

In this section we present results for the BER performance per subcarrier as a function of the subcarrier index $n$ for $m = 10$ and 2, for $\beta = 0.1$, 0.3 and 0.5, and $\alpha = 0.1$, 0.2, 0.3, 0.4 and 0.5. For $m = 10$ and $\beta = 0.1$ as shown in Fig. 3(a), it is evident that a compression ratio $\geq 20\%$ (i.e., $\alpha \geq 0.2$) cannot be supported using the 10-CAP at a BER level below the 7% forward error correction (FEC) limit of $3.8 \times 10^{-3}$ (shown as a dashed line in Figs. 3 and 4). This means that the data can be compressed by 10% with a BER lower than $10^{-4}$ for every subcarrier. It should be noted that the $\alpha = 0.1$ curve is not shown here, since no errors were found in the entire symbol stream. Thus, should the BER results for any value of $\alpha$ be lower than the $10^{-4}$ BER floor, the curve will be omitted from the figure. Interestingly, the BER values for the compression ratios of 20% and 30% are lower for $n = 1$ and 10. The reason for this is due to the fact that each of the subcarriers only interferes with its adjacent subcarrier, i.e., they only have one interfering subcarrier. However, for each subcarrier index $n = 2 - 9$, every individual subcarrier interferes with its two adjacent subcarriers, thus resulting in increased BER. Also shown in Fig. 3(a) are the constellation intensity maps for $n = 1$, 2 and 10 and for $\alpha = 0.3$. Considering 4-QAM with a compression of 10%, and $\beta = 0.1$, the resulting spectral efficiency is 2.02 b/s/Hz, which higher than 1.82 b/s/Hz for the uncompressed case with no degradation in the BER performance, thus giving a net spectral efficiency gain of around 11%.

In Fig. 3(b), for $\beta = 0.3$ and $m = 10$, it is possible to recover the signal with a BER $< 10^{-4}$ for $\alpha = 0.2$ and hence, the data is not plotted because every subcarrier is error free, and only plots for $\alpha > 0.3$ are shown. The reason for the improvement in the $\alpha = 0.2$ case over the previous results is due to the additional excess bandwidth introduced by the system in terms of $\beta$. Since increasing $\beta$ means the filter roll-off is slower and hence, the bandwidth is widened. Therefore, at the band-edges, the system can cope with additional interference due to a lower level of inter-modulation between subcarriers. Hence, a spectral efficiency of 1.92 b/s/Hz can be obtained over the original spectral efficiency of 1.54 b/s/Hz, giving a net spectral efficiency gain of 25%. Interestingly, this gain in spectral efficiency is not equal to the 20% compression rate and actually exceeds it, which may be differ from assumptions. The main reason stems from the fact the bandwidth $B$ is fixed at 3 MHz and the bit rate is given by $R_b = log_2(M)B/(1 + \beta)$. Hence, the spectral efficiency is clearly given by $\eta = R_b/B = log_2(M)/(1 + \beta)$. Introducing the compression factor $\alpha$, the spectral efficiency becomes

$$\eta = \frac{log_2(M)}{(1+\beta)(1-\alpha)} \quad (6)$$

similar to [17] but with the necessary additional roll-off term. For $\alpha = 0.3$, we observe that a link cannot be supported at a BER lower than the FEC target. Constellations are shown inset for $\alpha = 0.3$, $n = 4$ and 7.

For $\beta = 0.5$, $m = 10$, and $\alpha \leq 0.3$ the BER values are less than the FEC limit, whereas for $\alpha > 0.3$ the BER levels have increased well beyond the FEC limit. Also shown are the constellations for $n = 4$ and 6 for $\alpha = 0.3$, and $n = 9$ for $\alpha = 0.4$, respectively. Note that, the spectral efficiency is now increased to 1.91 b/s/Hz for $\alpha = 0.3$, compared to 1.33 b/s/Hz for the uncompressed case, thus representing an improvement of 44%, which is the highest relative improvement reported in this work. On the other hand, despite the fact that the largest compression factor is used, the spectral efficiency recorded for $\alpha = 0.3$, $m = 10$ and $\beta = 0.5$, represents the smallest spectral efficiency of the three cases demonstrated so far, by approximately 0.11 b/s/Hz. However, such a small difference in spectral efficiency may be a cost worth paying, since CAP systems with higher values of $\beta$ are more tolerant to timing jitter due to the wider horizontal opening of the pulse. This is advantageous, generally, but is especially relevant when considering that the received eye openings reduce with increasing carrier frequency [11]. However, this requires further investigation, which will be performed in future work.

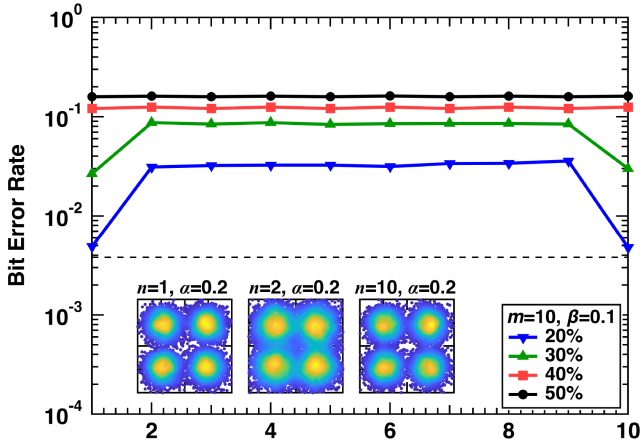

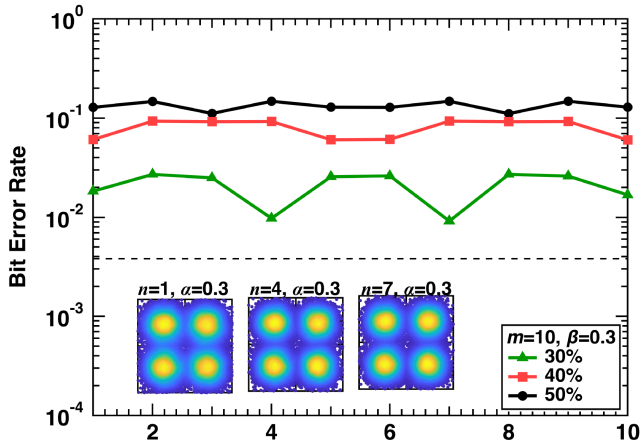

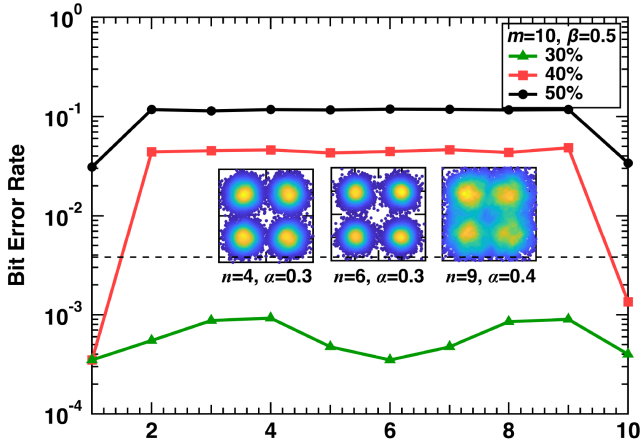

Fig. 3 BER plots for (a) $m = 10$ and $\beta = 0.1$, (b) $m = 10$ and $\beta = 0.3$ and (c) $m = 10$ and $\beta = 0.5$

A similar trend is observed for $m = 2$ and $\beta = 0.1$, as illustrated in Fig. 4(a), which shows the BER performance of the two-subcarrier system.

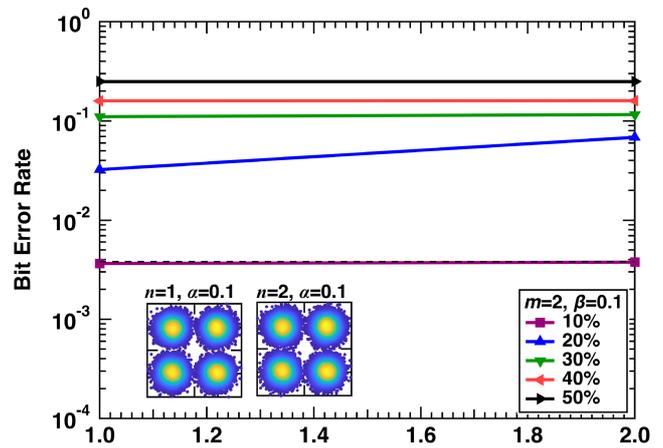

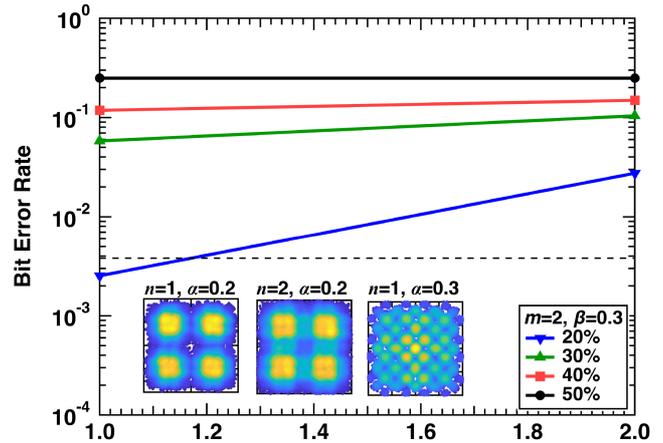

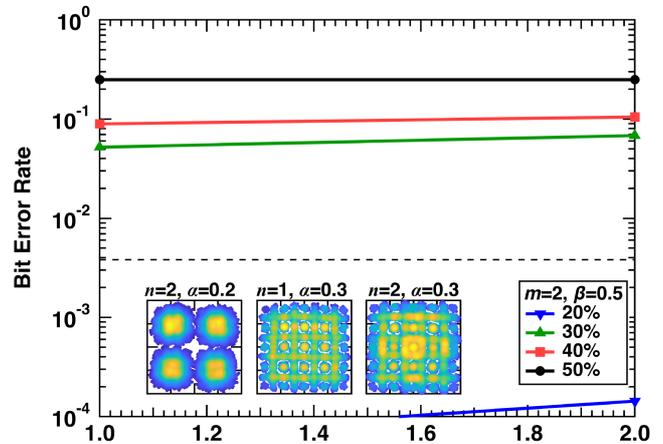

Fig. 4 BER plots for (a) $m = 2$ and $\beta = 0.1$, (b) $m = 2$ and $\beta = 0.3$ and (c) $m = 2$ and $\beta = 0.5$

It is clear that a link with $\alpha = 0.1$ can support a BER exactly equal to the FEC threshold of $3.8 \times 10^{-3}$. However, for $\alpha > 0.1$ the BER performance is increased beyond the FEC limit. This means that for $m = 2$ the maximum spectral efficiency available is exactly the same as the case for $\beta = 0.1$, and $m = 10$, i.e. a spectral efficiency of 2.02 b/s/Hz, representing a gain of 11% over the non-compressed case, albeit at a higher error rate (note that in the previous case the BER was < $10^{-4}$). This is an important result, because one must also consider the computational complexity when designing CAP systems, generally. The 2-CAP system uses 8 FIR filters (two pairs for in-phase/quadrature at both the transmitter and receiver, for two channels) while the 10-CAP system uses 40 filters, which represents a 500% increase in the comparative computational complexity. Also depicted in Fig. 4(a) are two constellation intensity maps for both subcarriers and $\alpha$ = 0.1.

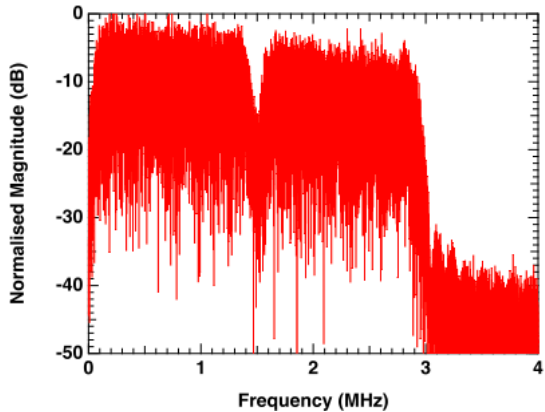

Fig. 5 The received electrical spectra of the 2-CAP signal showing the attenuation caused the low modulation bandwidth of the LED for $\alpha$ = 0 and $\beta$ = 0.1

In Fig. 4(b) show BER performance for $m$ = 2, $\beta$ = 0.3 and the same compression ratio as in Fig. 4(a). Note that, in this case it is possible to support $\alpha$ = 0.2. This is due to the wider subcarrier bandwidth, which has increased 5-fold to 1.5 MHz/subcarrier in comparison to $m$ = 10. However, wider bandwidths are more susceptible to the frequency attenuation experienced by the LED's low modulation bandwidth, which is illustrated using the electrical spectrum of the received signal in Fig. 5 ($\alpha$ = 0, $\beta$ = 0.1). Note, the increased attenuation for the second subcarrier. Hence, no additional spectral efficiency gain can be supported in spite of the higher roll-off factor, which is a disadvantage of a lower number of subcarriers. Constellations are shown inset for both $\alpha$ = 0.2 subcarriers (bottom) and $\alpha$ = 0.3, $n$ = 1. Interestingly, the constellations show the interference patterns due to intermodulation of the subcarriers and their deterministic nature. Since the received symbols are in fixed positions, the intermixing of subcarriers causes the complex multiplication of the two expected constellations, resulting in new received constellations with the observed interference patterns.

For the final case with $m$ = 2, $\beta$ = 0.5, and $\alpha$ = 0.3-0.5 the BER performance is illustrated in Fig. 4(c). The best BER performance (i.e., lower than the FEC limit of $3.8\times10^{-3}$) is achieved for $\alpha$ = 0.2 for $m$ = 2 in Fig. 4(a). Note that, the spectral efficiency is also less than $m$ = 10, thus supporting a total spectral efficiency of 1.67 b/s/Hz compared to 1.33 b/s/Hz for the uncompressed case, but at a much lower computational complexity.

Since no additional functionality is added at the receiver or transmitter in comparison to traditional orthogonal $m$-CAP, all of the results presented here demonstrate improved spectral efficiency, with a maximum gain of 44% ($m$ = 10, $\beta$ = 0.5) with no additional computational complexity in comparison to traditional 10-CAP system with no BER performance degradation

IV. CONCLUSIONS

In this paper, we propose and experimentally verify a novel modulation scheme called NM-CAP for increasing the spectral efficiency in bandlimited VLC systems. We demonstrated that for $m$ = 10, bandwidth compression up to 30% (i.e. 44% improvement in the spectral efficiency) can be supported without any BER performance degradation when compared to a conventional $m$-CAP. We also show that higher modulation order systems can support higher bandwidth savings than lower order ones. Moreover, the proposed novel scheme does not introduce additional computational complexity, which is a significant advantage. In future work we will focus on the optimization of NM-CAP system based on the measured signal-to-noise ratio values for individual subcarriers and bit- and power-loading algorithms.

ACKNOWLEDGEMENT

This work was supported by the UK EPSRC grant, EP/P006280/1: Multifunctional Polymer Light-Emitting Diodes with Visible Light Communications (MARVEL) and by GACR 17-17538S.